\documentclass{aa}

\usepackage{color}

\usepackage{amsmath}
\usepackage{natbib}
\usepackage{graphicx}
\bibpunct{(}{)}{;}{a}{}{,}
\usepackage{color}
\begin{document}

   \title{Inference of magnetic fields in inhomogeneous prominences}

   \subtitle{}

   \author{I. Mili\'{c}\inst{1,2}
   \and
   M. Faurobert\inst{3}
   \and
   O. Atanackovi\'{c}\inst{4}}

   \institute{Max-Planck-Institut f\"{u}r Sonnersystemforschung, Justus-von-Liebig-Weg 3, 37075 G\"{o}ttingen, Germany\\
   \email{milic@mps.mpg.de}
         \and
         Astronomical observatory Belgrade, Volgina 7, 11060 Belgrade, Serbia\\
          \and
          UMR 7293 J.L. Lagrange Laboratory, Universit\'{e} de Nice Sophia Antipolis, CNRS, Observatoire de la C\^{o}te d'Azur, Campus Valrose, 06108 Nice, France\\
              \email{marianne.faurobert@oca.eu}
           \and
           Department of astronomy, Faculty of Mathematics, University of Belgrade, Studentski Trg 16, 11000 Belgrade, Serbia\\
           \email{olga@matf.bg.ac.rs}
             }

   \date{}
   \titlerunning{The inference of magnetic fields in inhomogeneous prominences}
   \authorrunning{Mili\'{c} et al.}


  \abstract
   {Most of the  quantitative information about the magnetic field vector in solar prominences comes from the analysis of the Hanle effect acting on lines formed by scattering. As these lines can be of non-negligible optical thickness, it is of interest to study the line formation process further.}
  {We investigate the multidimensional effects on the interpretation of spectropolarimetric observations, particularly on the inference of the magnetic field vector. We do this by analyzing the differences between multidimensional models, which involve fully self-consistent radiative transfer computations in the presence of spatial inhomogeneities and velocity fields, and those which rely on simple one-dimensional geometry.}
   {We study the formation of a prototype line in ad hoc inhomogeneous, isothermal 2D prominence models. We solve the NLTE polarized line formation problem in the presence of a large-scale oriented magnetic field.  The resulting polarized line profiles are then interpreted (i.e. inverted) assuming a simple 1D slab model. }
   {We find that differences between input and the inferred magnetic field vector are non-negligible. Namely, we almost universally find that the inferred field is weaker and more horizontal than the input field.} 
   {Spatial inhomogeneities and radiative transfer have a strong effect on scattering line polarization in the optically thick lines. In real-life situations, ignoring these effects could lead to a serious misinterpretation of spectropolarimetric observations of chromospheric objects such as prominences.}

   \keywords{Line: formation; Radiative transfer; Polarization; Sun: filaments, prominences}

   \maketitle


\section{Introduction}

Magnetic fields play a crucial role in the physics of the outer solar atmosphere. Among other phenomena, magnetic effects are the origin of the formation and evolution of various chromospheric structures such as solar prominences (filaments) and spicules. To understand these objects it is essential to compare the magnetic field vector inferred from the observations with the magnetic field resulting from theoretical studies. The main sources of quantitative information on the magnetic field pervading these objects are spectral lines sensitive to the   Hanle and Zeeman effects. The 10830$\,\rm{\AA}$ and D3 (5876$\,\rm{\AA}$) lines of  helium are among the most frequently used spectral lines for such an analysis. To interpret the observations performed in these lines, different diagnostic tools have been devised, ranging from simple polarization diagrams \citep[e.g.][]{LL04} to modern inversion codes such as HAZEL \citep{HAZEL} and HELIX \citep{HELIXX}. 

The magnetic field is not an observable quantity; it is indirectly inferred from the observed Stokes vector. The interpretation of any spectropolarimetric observation is thus done assuming a specific generative model. The observed data are values of Stokes parameters for a given set of wavelengths and their errors are usually Gaussian and follow from the signal-to-noise ratio. The generative model is inevitably simple and relies either on what is known as a  single-scattering approximation or on a 1D slab model, which can also involve some elementary 1D radiative transfer. Although these generative models are oversimplified when compared to the ``real'' Sun, we are, in a way, condemned to use them because more sophisticated radiative transfer calculations are computationally very demanding and inversion techniques based on self-consistent multidimensional polarized NLTE radiative transfer are still unfeasible.

In this paper we investigate the bias that emerges when a simple 1D model is used to interpret the observations that are, in fact, generated in a more complex (and more realistic) multidimensional, dynamic, and inhomogeneous medium. We study the polarized spectral line formation process in a 2D Cartesian slab standing above the solar surface and illuminated by the solar radiation. The slab is inhomogeneous in density and pervaded by different velocity fields and by a uniform vector magnetic field. We consider a prototype line, formed in the two-level approximation (normal Zeeman triplet), under the assumption of complete frequency redistribution. These assumptions simplify the atomic physics involved and allow us to focus on ``macroscopic'' effects on the line formation process. We compute synthetic spectra from these models and then use a 1D scheme to ``invert'' these synthetic observations. We discuss the differences between the input parameters and the inferred ones and point to possible systematic errors and degeneracies in the process of the inference of the magnetic field vector in such objects.


\section{Prominence model and computation of the spectra}

\begin{figure}
\includegraphics[width = 0.5\textwidth]{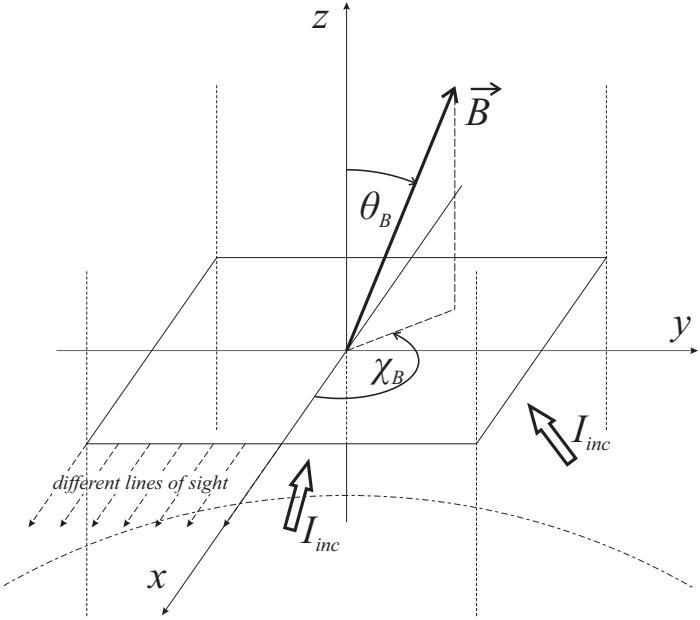}
\caption{The geometry of the problem. The prominence is represented by a vertical slab that is finite and inhomogeneous in the $xy$ plane and infinite and homogeneous along the $z$-axis. The slab is illuminated from the sides by the limb-darkened solar radiation. The line of sight corresponds to $\theta = \pi/2$, $\varphi = 0$.}
\label{fig_geometry}
\end{figure}

We base the geometry of our multidimensional prominence model on the work of \citet{Heinzel_Prom2D}. They were the first to introduce multidimensional radiative transfer to the computations of prominence spectra \citep[see also][for subsequent applications and comparison with the observations]{Heinzel_Prom2DII, Gunar12}. The prominence is modeled as a 2D Cartesian slab, finite and inhomogeneous in $x$ and $y$ directions and infinite in $z$ direction ($z$  is atmospheric normal, see Fig.\,\ref{fig_geometry}). The direction of propagation of radiation and the magnetic field orientation are given in this referent system. The slab is pervaded with a velocity field which have components along all three directions. It is illuminated by the limb-darkened solar radiation coming from the sides. This illumination provides boundary conditions for the solution of the NLTE polarized line formation problem described below. We consider only isothermal prominences (there is no prominence-to-corona transition region). As we will see from the following subsection, the radiative transfer problem is uniquely specified by the opacity distribution, the height of the slab above the solar limb, the incident radiation, and the atomic model.


\subsection{Synthesis of emergent Stokes vector}
\label{RT}

We assume that the line is formed by resonant scattering in the framework of a two-level atom model under the assumption of complete frequency redistribution (CRD). We study a prototype spectral line with rest-frame line center wavelength equal to $10830\,{\rm \AA}$ and Doppler width equal to $0.2\,\rm{\AA}$, which corresponds to the random velocity of $\approx5$\,km/s. The upper level inverse radiative lifetime of the transition is $A_{\rm{ul}} = 1\times10^7\,\rm{s}^{-1}$. We assume that the line is broadened only by the Doppler effect due to thermal motion of the scattering atoms, which leads to a Gaussian shape of the line-absorption profile. The Land\'{e} factor of the transition is equal to 1.5. The intrinsic line polarizability (see below) is set to $0.3$. The values of the line parameters mentioned above are chosen so that the polarization rates and magnetic field sensitivity correspond to the helium\,10830 line, which is often used for prominence magnetic field diagnostics \citep[e.g.][]{David14}. 


In spectral lines formed out of local thermodynamic equilibrium (LTE) approximations, a part of the linear polarization (i.e. Stokes $Q$ and $U$) arises from the scattering of anisotropic incident radiation, which gives rise to population imbalances in the upper level of the transition (in the case we are considering, the lower level is unpolarized). This polarization is then further modified by the presence of the magnetic field via the Hanle effect \citep[e.g.][]{LL04}. The analysis of the observed Stokes spectrum then, in principle, allows us to find the magnetic field vector.

The equations which govern the transfer of polarized radiation, under the above assumptions and with no background continuum, are given by
\begin{equation}
\frac{d\hat{I}(x,y,\theta,\varphi,\lambda)}{d\tau} = \phi(\lambda) \left [ \hat{I}(x,y,\theta,\varphi,\lambda) - \hat{S}(x,y,\theta,\varphi) \right ]
\label{RTE}
\end{equation}
and the equation of statistical equilibrium for a two-level atom,
\begin{align}
\hat{S}(x,y,\theta,\varphi) = & \hat{W} \int_{-\infty}^{\infty} \phi(\lambda) d \lambda \oint \frac{\sin{\theta} d\theta d\varphi}{4\pi} \times \nonumber \\
& \hat{P}(\theta, \varphi, \theta', \varphi') \hat{I}(x, y, \theta, \varphi, \lambda).
\label{SE}
\end{align}
Here, $\hat{I}$ is the polarized intensity (i.e. Stokes vector), $\hat{I} = (I,Q,U,V)^T$; $\hat{S}$ is the polarized source function, i.e. the ratio of emissivity and opacity; and $d \tau = \chi_{\rm{L}} ds$ (where $ds$ is the elementary geometrical path and $\chi_{\rm{L}}$ is so called line-integrated opacity) is the optical path along the ray. The matrix $\hat{W}$ contains contributions from different (de)polarizing mechanisms. When depolarizing collisions are absent, $\hat{W} = \hat{W}_H \hat{W}_2$, where $\hat{W}_H$ is the magnetic matrix \citep{Frisch07} and $W_2$ is the  intrinsic line polarizability, which  in this case is equal to $\rm{diag}(1, 0.3, 0.3, 0)$.  The line-absorption profile $\phi(\lambda)$ is taken to have Gaussian shape. Finally $\hat{P}$ is the scattering matrix which describes the angular coupling between the incoming and outgoing photons \citep[see e.g.][]{Faurobert87, Bommier97}. In the case of resonant line scattering, the matrix $\hat{P}$ corresponds to the Raighley's scattering matrix, while in the multilevel atom case  a different approach must be used, namely the density matrix formalism described in \citet{LL04}. We solve the coupled equations \ref{RTE} and \ref{SE} by employing the real reduced intensity formalism described in \citet{Frisch07} and generalized to multidimensional geometries by \citet{AnushaI}, and by performing Jacobi iterations to solve the coupled equations in the reduced basis. 

The formalism briefly described above allows us to compute the $I$, $Q$, and $U$ components of the Stokes vector created in the scattering processes and modified by the magnetic field (Hanle effect). The $V$ component, however, is due to the Zeeman effect. In the particular case of a simple prominence model, where the magnetic field is assumed to be weak and constant through the medium, Stokes V is given by the so-called weak field approximation \citep{LL04}
\begin{equation}
V(\lambda) = -4.39\times10^{-13} \lambda^2 g B_{||} \frac{d I}{d\lambda}
,\end{equation}
where $g$ is the Land\'{e} factor and $B_{||}$ is the line-of-sight component of the magnetic field vector. In this expression, the wavelength is given in angstrom and the magnetic field strength in gauss. The Zeeman effect also modifies Stokes $Q$ and $U$, but in this particular case of weak magnetic fields these effects are negligible when compared to the scattering polarization and Hanle effect. 

After the self-consistent solution for the polarized source function has been found,  it is possible to perform  formal solutions in the desired directions and at the desired wavelengths to synthesize the emergent Stokes vector. For a given atomic model, the formation of the emergent spectra is completely determined by the height of the slab above the solar limb $H$ (which we take to be equal to $20$ arcseconds, approximately $15\,$Mm), by the opacity and velocity distributions, and by the incident radiation. To obtain the radiation field that illuminates the slab we use the limb-darkened continuum radiation corresponding to the solar limb darkening law in the near-infrared. In this paper we confine ourselves to the case where the observed object is at the solar limb, so we observe the radiation emerging from the object in the direction $\theta = \pi/2$ and $\varphi =0$ (see Fig.\,\ref{fig_geometry}).

\subsection{Interpretation of the observed Stokes vector}

Scattering polarization in general, carries two types of information,  on the radiation anisotropy inside the object and  on the magnetic field vector. We discuss these two aspects in a bit more detail: in the absence of the magnetic field and any azimuthal anisotropies Stokes $Q$ depends solely on the vertical anisotropy. The existence of non-zero Stokes U indicates either the presence of a magnetic field or of  another symmetry-breaking effect, such as the azimuthal anisotropy of the radiation field. ssume an optically thin ensemble of atoms situated above the solar surface, and scattering of the incident radiation. In this  single-scattering scenario, Stokes $Q$ is non-zero because the scattering atoms are illuminated by a cone of limb-darkened radiation, while Stokes $U$ is zero because the radiation is axisymmetric with respect to the $z$-axis. The magnetic field rotates the plane of polarization and creates, in the general case, non-zero $Q$ and $U$. This, together with the measurement of Stokes V, allows magnetic field diagnostics in optically thin plasma \citep[e.g.][]{LL04, HAZEL}. However, observations and analysis of prominence spectra show that they are of moderate optical thickness in certain diagnostically important lines \citep[e.g. He\,10830,\,][]{David14}. 

Scattering in optically thick medium leads to various radiative transfer effects. In particular, if the medium is inhomogeneous, we are to expect non-zero Stokes $U$ even in the absence of a magnetic field. The presence of a magnetic field modifies both $Q$ and $U$. It is important to understand these effects and take them into account as solar prominences, filaments, and spicules exhibit complicated (i.e. inhomogeneous) and dynamic structures. However, all interpretations so far have used simple generative models to interpret the observations. By definition, these models (either the single-scattering or simplified 1D slab) do not account for these multidimensional effects. This means that if multidimensional effects are present in the real-life prominences, spectral diagnostics relying on a model that neglects them would be fundamentally wrong. 

Our aim in the remainder of the paper is to show the importance of the effects of inhomogeneities, velocity fields, and radiative transfer on polarized line formation in solar prominences. We  start with a very simple example.

\section{Illustrative examples}

\subsection{Homogeneous 2D slab}
\label{ex_1}

As a reference case, we first compute and plot emergent polarization from a homogeneous, static and non-magnetized slab which is finite in the $xy$ plane. The slab has an optical depth along $x$ (line of sight) equal to unity and an optical depth along $y$ equal to 5. As this is an illustrative example, and the prominence is homogeneous, we do not relate the optical depth to its geometrical dimensions.

\begin{figure*}
\includegraphics[width = \textwidth]{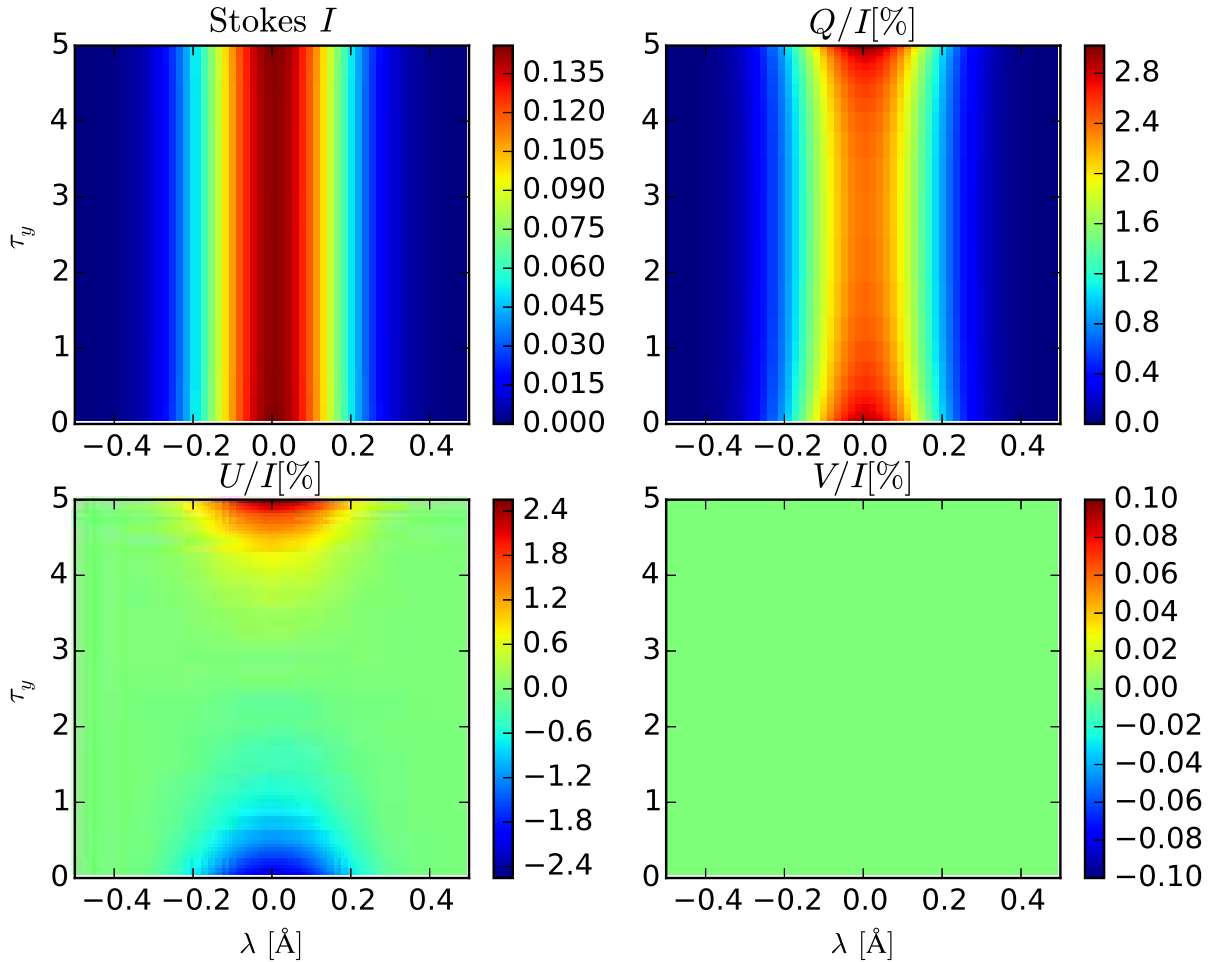}
\caption{Synthetic polarized spectra emerging from a homogeneous 2D slab, as described in section\,\ref{ex_1}.}
\label{example_1}
\end{figure*}

Figure\,\ref{example_1} shows the spatial distribution of the Stokes profiles. These density plots can be understood as synthetic spectra where the light has been dispersed along the horizontal axis, while the vertical axis shows the spatial variation of the Stokes profiles along the spectrograph slit. In this case, we show the spatial distribution along the $y$-axis (see Fig.\,\ref{fig_geometry}). This would correspond to a slit positioned parallel to the solar limb. Stokes $Q$, $U$, and $V$ in every pixel are normalized with respect to the maximum intensity in the corresponding pixel.

The $I$ component is identical to what would be obtained with a 1D model. Stokes $V$ is zero because there is no magnetic field in the slab. The $Q/I$ and $U/I$ components, however, are not constant along the slab. The most  prominent effect is a strong polarization in Stokes $U/I$ because of the breaking of the left/right symmetry close to the edges of the slab. Namely, outside the ``cone'' of radiation that illuminates the slab, the incoming radiation is zero, while the emergent radiation is not. This breaks the azimuthal (left-right) symmetry and results (in this case) in a significant Stokes $U/I$ signal. Moving toward the center of the slab, the radiation becomes more and more axisymmetric and $U$ tends to zero. This also explains the increase of Stokes $Q/I$ close to the edges. The increase in the polarization is not as prominent, as up-down asymmetry (anisotropy of distribution of the radiation in $\theta$) exists everywhere in the slab. Neither of these two effects exists in 1D models because they are infinite both in $y$ and $z$ and it is impossible to break the left-right symmetry in the absence of a magnetic field. We note that this last statement is strictly valid only for this viewing angle ($\theta = \pi/2$,$\phi =0$). To see the effects of different viewing geometry, see  e.g.  \citet{Bommier_89}. 

Realistic astrophysical objects, however, do not have such sharply defined boundaries and thus we are unlikely to see the ``edge'' effects described above in real-life objects. Another effect that can create polarization in Stokes $U$ are inhomogeneities in the object. We now present a toy model of prominence which demonstrates these effects.

\subsection{Two-dimensional slab with horizontal inhomogeneities: Magnetic field-free case}
\label{ex_3}

For the following example we study a toy model of a multi-threaded prominence. The presence of threads is well established now \citep[for a detailed reviews on prominence structure, spectrum formation and oscillations see][]{Prom_review_I, Prom_review_II}, and radiative transfer computations which incorporate such structures have been done \citep{Leger09, Gunar12}. None of these, however, involve any computations of scattering polarization in spectral lines, which is a pity because polarization is supposed to be an excellent probe of the inhomogeneities in the medium. We illustrate this in the following example. 

We consider an inhomogeneous, isothermal prominence which consists of $N=50$ overdensities which have a Gaussian shape in the $xy$ plane. That is, our opacity (density) structure has the  form
\begin{equation}
 \chi(x,y,\lambda) = \chi_0 \sum_i^{N} e^{(\left[-(x-x_{0,i})^2-(y-y_{0,i})^2 \right] / \sigma^2)} \phi(\lambda)   
,\end{equation}
where $x_{0,i} $ and $y_{0,i}$ are the center coordinates of the i-th thread and $\chi_0$ is constant, chosen so that the spatially averaged, line-integrated optical depth along the line of sight is equal to unity. That is,
\begin{equation}
 \chi_0 = \frac{1}{y_{\rm total}}\int_0^{x_{\rm total}}\int_0^{y_{\rm total}} \int_{0}^{\infty} \chi(x,y,\lambda) dx dy d\lambda = 1,
\end{equation}
where $\sigma$ is the half-width of the threads; in the following examples we take $\sigma =150\rm{km}$. The dimensions of the prominence are $x_{\rm total} = 2000\,{\rm km}$ and $y_{\rm total} = 16000\,{\rm km}$ (approx $22$ arcseconds in the plane of the sky). The prominence height and the  atomic model parameters are the same as in the previous section. We first discuss the emergent spectrum obtained in the absence of  magnetic field and we leave the magnetized case and its interpretation for the next section.

\begin{figure*}
\includegraphics[width = \textwidth]{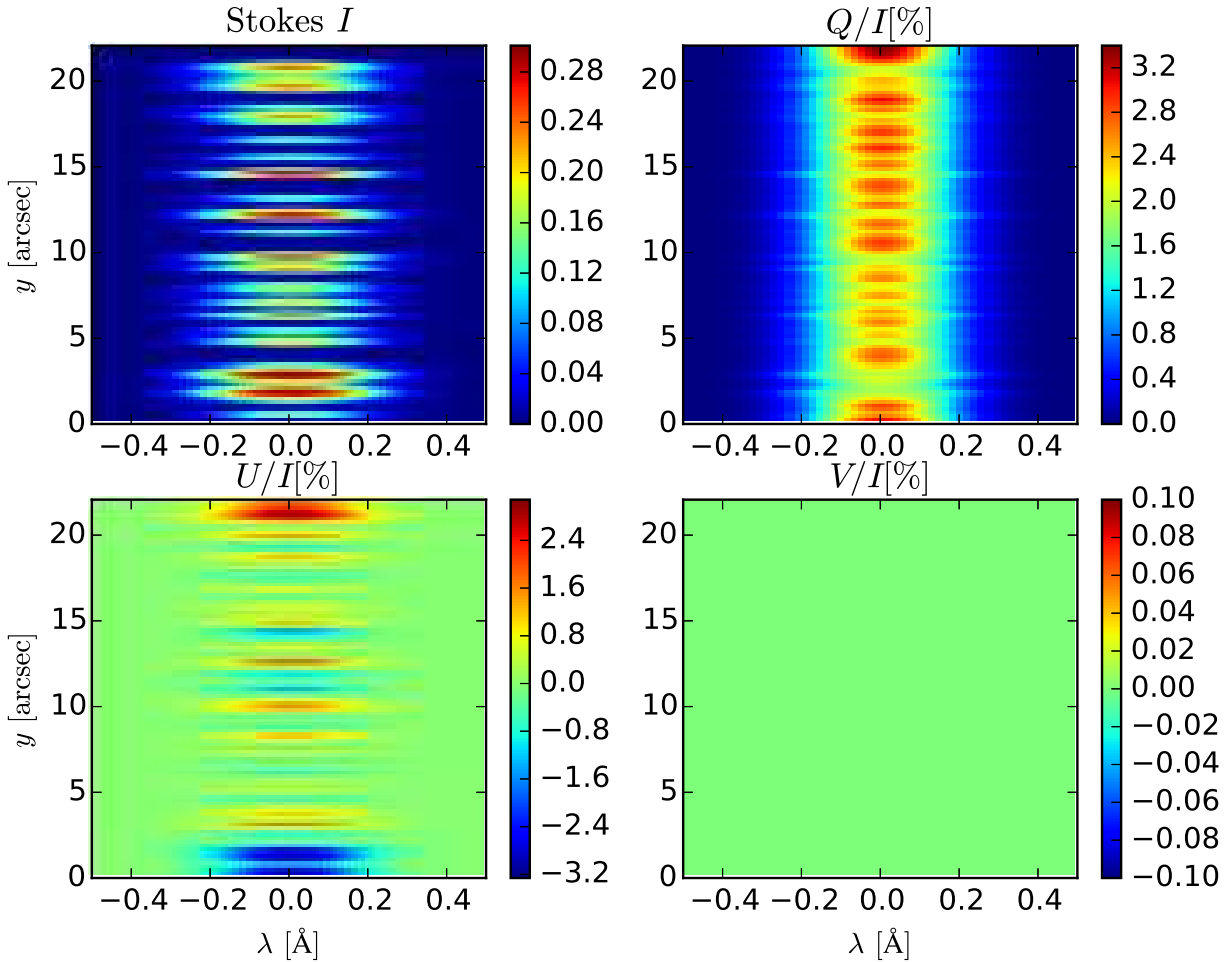}
\caption{Synthetic polarized spectra emergent from an inhomogeneous slab in absence of the magnetic field. We note the non-zero Stokes $U$, which is the consequence of azimuthal anisotropy.}
\label{example_3}
\end{figure*}

The emergent synthetic Stokes spectra are shown in Fig.\,\ref{example_3}. First three Stokes components now show fluctuations over the slit. The intensity follows the distribution of the opacity. As the prominence is still optically thin, in the first approximation we can assume that the source function for Stokes $I$, $S_{I}$, is constant and equal to the incident mean intensity. In this case, the formal solution of the radiative transfer equation yields
\begin{equation}
 I(\lambda) = S_I (1 - e^{-\tau_{\lambda}}), \nonumber
\end{equation}
where $\tau_{\lambda}$ is the total optical depth of the prominence along the line of sight at the wavelength $\lambda$. As the opacity per particle is assumed to be constant in the prominence, the optical depth is proportional to the column density along the line of sight.

The distribution of Stokes $Q/I$ is anti-correlated with Stokes $I$. This can be understood from the following argument: a more opaque medium results in a more highly isotropic radiation inside owing to multiple scatterings and hence in a smaller $Q$ component of linear polarization. We note that this effect  can be explained qualitatively by 1D models as well. The spatial variations of Stokes $U$ are, however, a purely multidimensional effect. They arise because the variations in opacity directly result in the axial asymmetry of the radiation field which, in turn, results in a Stokes $U$ signal. This effect  has already been studied by e.g. \citet{TBSr07} and \citet{StepanLyAlpha15} for the inhomogeneities in the solar atmosphere. In the particular case of  prominences, this effect is equally important as the physical conditions in prominences are very similar to those in the chromosphere. There are two main reasons why these effects might influence our inference from spectropolarimetric observations. First,  prominences are very dilute and the mean optical path of photons could easily be larger than the spatial resolution of the telescope. This means that the coupling of different regions in the prominence due to the scattering processes could actually be observed. Second,  in simple prominence magnetic field diagnostics the presence of Stokes $U$ is interpreted as the presence of a magnetic field.  Spurious $U$ signals due to multidimensional effects could easily lead to a  misinterpretation of prominence spectra and misdiagnosis of the magnetic field. 


\section{The magnetized case}

The discussion above illustrates the appearance of a ``non-magnetic'' rotation of the polarization plane. That is, we have explained how multidimensional effects can create polarization in Stokes $U$ in the absence of a magnetic field. We are, however, practically certain that magnetic fields are present in solar prominences. It is of interest, then, to repeat the computations of the previous section, but now taking into account the Hanle and Zeeman effects of a given magnetic field. We use the same opacity distribution and prominence model as in the previous case, but we assume that the prominence is pervaded with a homogeneous magnetic field with magnitude 5 gauss and orientation ($\theta_{\rm B} = 75^{\rm o}$, $\phi_{\rm B} = 50^{\rm o}$). 

\begin{figure*}
\includegraphics[width = \textwidth]{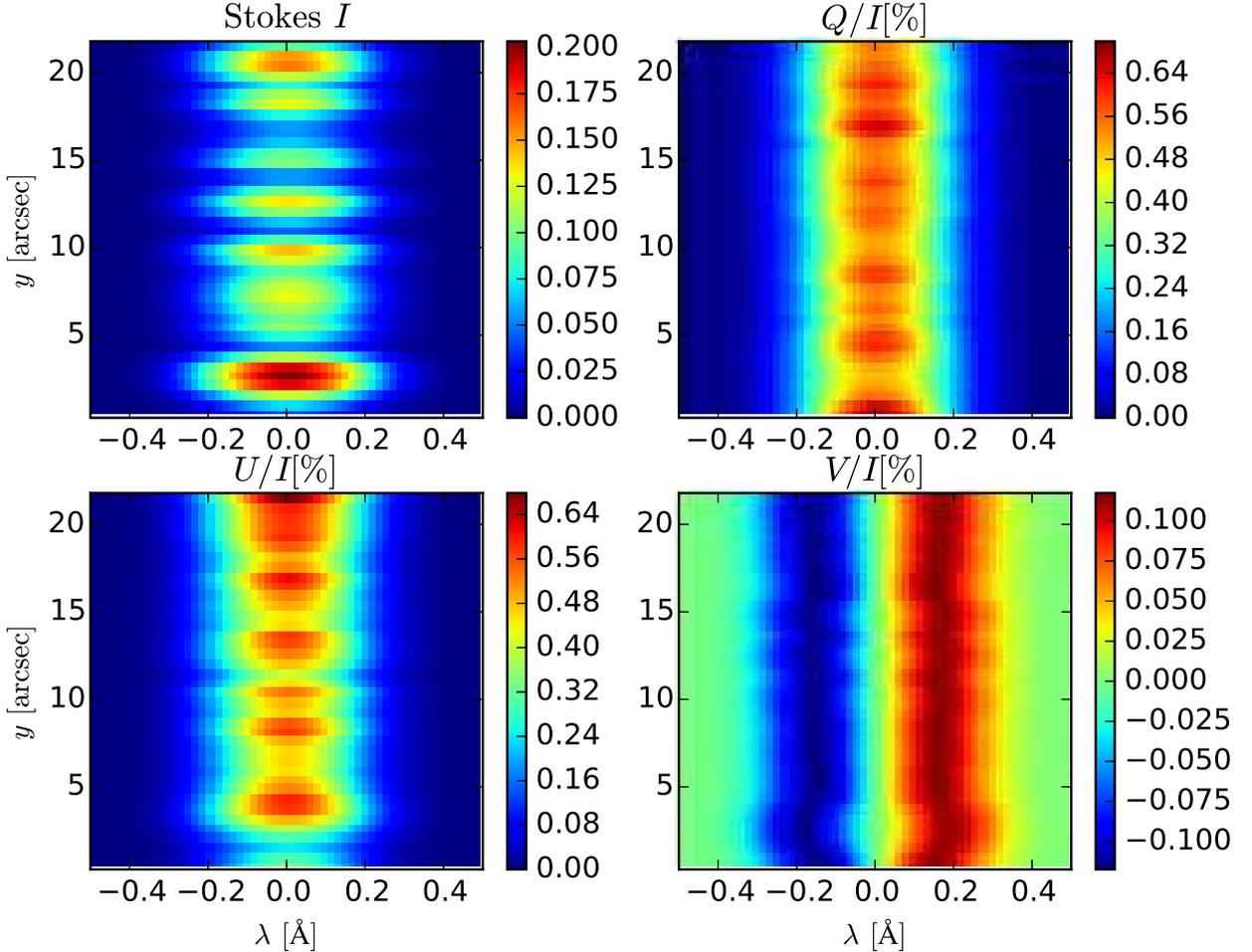}
\caption{Same as Fig.\,\ref{example_4}, but the slab is now pervaded with magnetic field and spatial degradation has been applied. See the text for details.}
\label{example_4}
\end{figure*}

To bring our computations closer to actual observations we also convolve the spatial distribution of the Stokes profiles with a Gaussian point spread function with half-width equal to $0.4$ arcseconds (which is a crude approximation to a 1 arcsecond wide Airy disk) to simulate the smearing effects of the limited angular resolution of a ground-based instrument. Afterward, we bin our data in $0.5$ arcsecond wide bins. The resulting spectra are shown in Fig.\,\ref{example_4}. 

Both Stokes $Q$ and $U$ are non-zero everywhere in the prominence. The interpretation of their magnitudes, shapes, and spatial distribution is, however, not trivial 
because we are faced with the coupled effects of multidimensional radiative transfer and the Hanle effect. We note that Stokes $V$ is given by the weak field approximation and that, in principle, it is directly proportional to the component of the magnetic field parallel to the line of sight. 

In the next step, we devise a simple 1D inversion scheme and apply it to the synthetic data shown in Fig.\,\ref{example_4}.

\subsection{Interpretation of the synthetic data}
\label{inversion_model}

We base our interpretation scheme on a 1D slab model, as in e.g.  \citet{HAZEL}. We parametrize the slab (i.e. the generative model), with the following parameters:
\begin{itemize}
 \item Line-integrated optical depth ($\tau$) of the slab along the line of sight. This parameter influences the shape and the magnitude of the emergent Stokes vector. If the radiative transfer effects inside the slab are accounted for, it also influences the anisotropy of the radiation. For example, in the magnetic field-free case, an increase in $\tau$ would result in a decrease in the vertical anisotropy and thus in a smaller signal in Stokes $Q$;
 \item Line-center position ($\lambda_0$). This parameter accounts for the systematic velocity of the slab as a whole. In practice, it only shifts the profile left and right;
 \item Doppler half-width of the line ($\Delta\lambda_{\rm D}$) given in terms of the turbulent velocity $v_t$, which is a combination of thermal (Maxwellian) and microturbulent velocities;
 \item Magnetic field vector, given by its magnitude and orientation ($B$, $\theta_B$, $\phi_B$).
\end{itemize}
To interpret the data we need to describe the mapping $\hat{M} \rightarrow \hat{I}$, where $\hat{M} = (\tau, \lambda_0, \Delta\lambda_D, B, \theta_B, \phi_B)$ and $\hat{I}$ is the emergent Stokes vector. This mapping is, of course, given by the radiative transfer processes. That is, we need to find a self-consistent solution of the equations of radiative transfer and statistical equilibrium (Eqs.\,\ref{RTE} and \ref{SE}) in the presence of the magnetic field, this time in 1D. We first solve the unpolarized problem to find the vertical anisotropy of the radiation, i.e. the first two components of reduced mean intensity \citep{AnushaI}. Then, we account for the effect of the magnetic fields to compute the emergent Stokes vector. In this procedure we assume that the polarized source function is dominated by the angular distribution of Stokes $I$, which is taken to be the same as in the unpolarized case. This is similar to the approach taken by \citet{HAZEL} in their forward modeling of the Hanle and Zeeman effect in the helium atom. Strictly speaking, in optically thick plasma, all Stokes parameters significantly influence radiation field anisotropy. This effect requires detailed calculations and can prove to be important \citep[see e.g.][]{JTB01}. We omit it here for simplicity but also because it is not accounted for in current Hanle inversions.

After the generative model is fully described and we are able to compute $\hat{I}$ given $\hat{M}$, we need to specify the merit function. In general, we are interested in the shape of the probability density function of the parameters given the data, that is $p(\hat{M}|D)$, where $D$ stands for ``Data'' (the observed Stokes vector) with their corresponding uncertainties (errors). Bayes' theorem states that
\begin{equation}
 p(\hat{M}|D) = \frac{p(D|\hat{M})p(\hat{M})}{p(D)}
,\end{equation}
where the left-hand side is known as the  {a posteriori} probability, $p(D|\hat{M})$ is the likelihood, $p(\hat{M})$ is the {a priori} probability, and $p(D)$ is known as the evidence. Finding the maximum of the {a posteriori} probability (MAP) in the hyperspace of parameters $\hat{M}$ and, more generally, the shape of $p(\hat{M}|D)$ is the aim of the  Bayesian methods \citep{McKay}. These have been seldom used in solar physics, but see \citet{AAR07} and \citet{AAR11} for different insightful examples. 

Under the assumption of Gaussian uncertainties only on the independent variable (Stokes vector), the likelihood function has the  form
\begin{equation}
 p(D|\hat{M}) \propto e^{-\frac{1}{2}\chi^2(\hat{M}, D)}
,\end{equation}
where
\begin{equation}
 \chi^2(\hat{M}, D) = \sum_i \sum_s \left(I_{si}^{\rm O} - I_{si}^{\rm C}(\hat{M}) \right)^2 w_{si},
\end{equation}
where the indices $s$ and $i$ refer to the different Stokes components and different wavelength points and the superscripts O and C refer to the observed and computed Stokes vector. 
The weights $w_{si}$ in principle follow from the noise of the individual points, which is  inferred from the signal-to-noise ratio of the observations. It is important to note that the evidence ($p(D)$) does not depend on the parameters. Now, in the case where all the parameters $\hat{M}$ are equally probable (the so-called uniform priors), finding the maximum of the posterior probability reduces to finding the maximum of the likelihood which is, under these assumptions, nothing else but the minimization of $\chi^2$.

We impose some simple priors for our parameters. The optical depth of the slab, the thermal velocity, and the magnetic field strength have to be positive. The priors for these parameters are thus step functions. We also impose that the azimuth of the magnetic field must be between zero and $90^o$. This is mostly to avoid ambiguities as we know the ``real'' magnetic field and just want to see how much the inversion misses owing to the inadequate model. Finally, we restrict the inclination to the range $(0, 180^o)$ and use the prior
\begin{equation}
 p(\theta_B) \propto \sin\theta_B
,\end{equation}
which corresponds to an isotropic magnetic field distribution.

There are a number of ways to find MAP (or, when priors are not important, to minimize $\chi^2$). Here we opt for the simple but robust Affine-invariant Markov chain Monte Carlo method \citep[MCMC, see][]{AiMCMC}. The implementation is done in python, under the ``emcee'' package \citep{Hammertime!}. This approach allows us not only to find the maximum (hopefully global)  of the {a posteriori} probability, but also to sample around it, analyze the marginalized probability density distributions of individual parameters (to reliably evaluate the parameter uncertainties) or parameter combinations (to study the degeneracies between the parameters), and in general to probe the shape of the $p(\hat{M}|D)$ function in the hyperspace of parameters. Before inverting the synthetic data, we verify the fitting procedure on a single scattering case and find that it recovers the  original magnetic field well.


\subsection{Inversion results: Noise-free case}

We now invert (i.e. fit, find MAP) each of synthetic pixels given in Fig.\,\ref{example_4} using the model presented in the previous section. We note that, in principle, the likelihood, and thus the 
 {a posteriori} probability density functions are defined for given values of the measurement uncertainties. To simulate observational data it is necessary to add (in the simplest approximation) a Gaussian noise of a given level to these observations. We first invert the synthetic observations adding a very low level of wavelength independent noise equal to $10^{-6}$ times Stokes $I$ in the disk center.

We expect very good fits of Stokes $I$ line profiles as the line is everywhere close to being optically thin and because inhomogeneities do not influence the scalar source function (and thus intensity) as much as the polarized one. It is  also expected that the fitting procedure (regardless of the specific implementation) will easily fit the intensity and the circular polarization and will then try to find the best possible fit for the linear polarization. Now, because the linear polarization is generated via multidimensional effects, this might not be the case. The fitting procedure might (erroneously) offset a good fit in Stokes $V$ with fits in Stokes $Q$ and $U$  

\begin{figure}
\includegraphics[width = 0.5\textwidth]{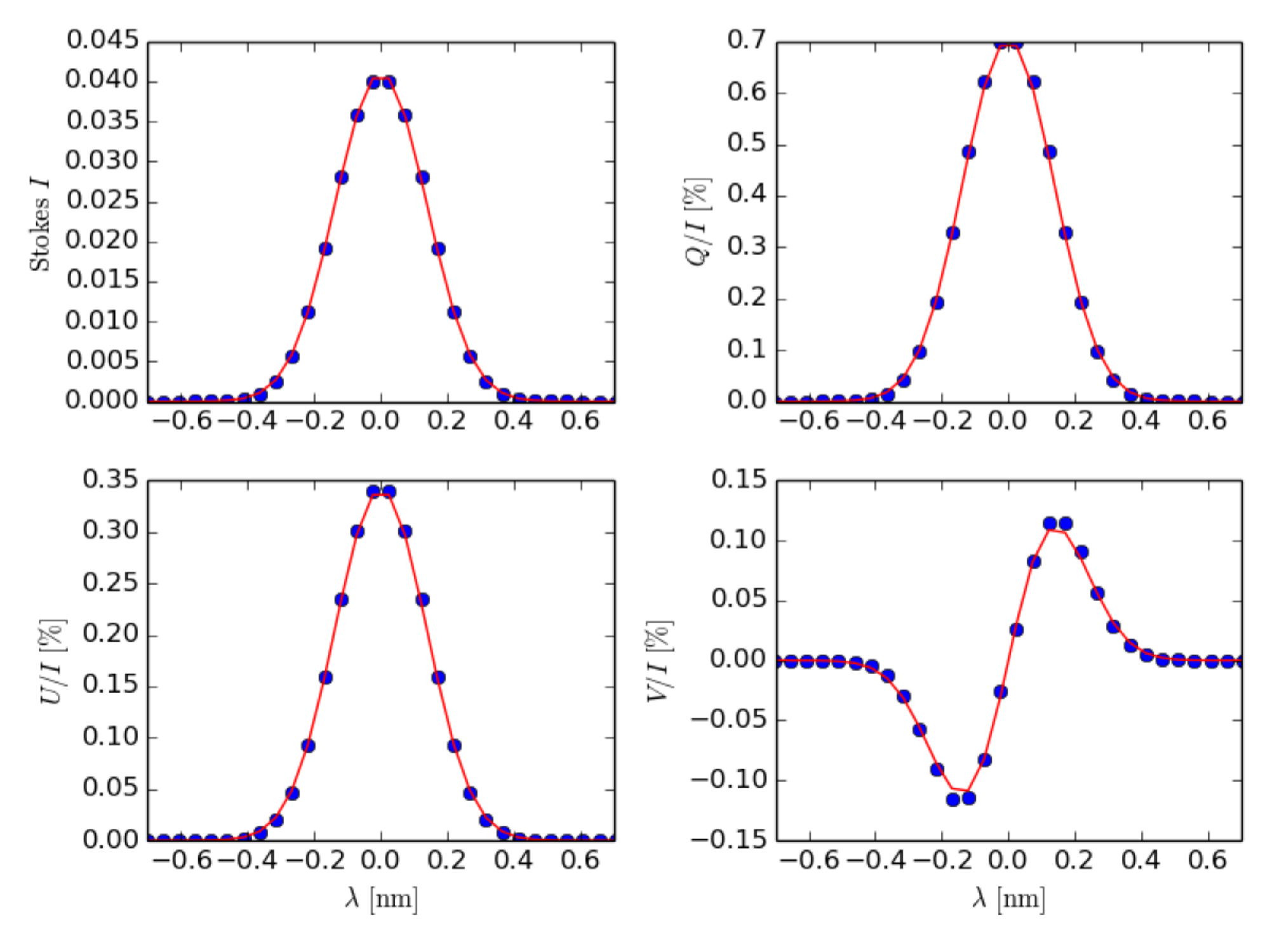}
\caption{Stokes spectrum emerging from the location with $y=0$'', and the best fit solution.}
\label{spectrum_0}
\end{figure}

First we analyze the profiles emerging on the edge of the slab. Fig.\,\ref{spectrum_0} shows the best fit for the first pixel (i.e. the one with $y=0$). Interestingly, the fit for all the Stokes components is quite good. However, the inferred parameters are somewhat different from the input values. The inferred magnetic field is weaker ($3.9\pm0.1\,$gauss), and more inclined and directed toward the observer ($87^o,39^o$).

\begin{figure}
\includegraphics[width = 0.5\textwidth]{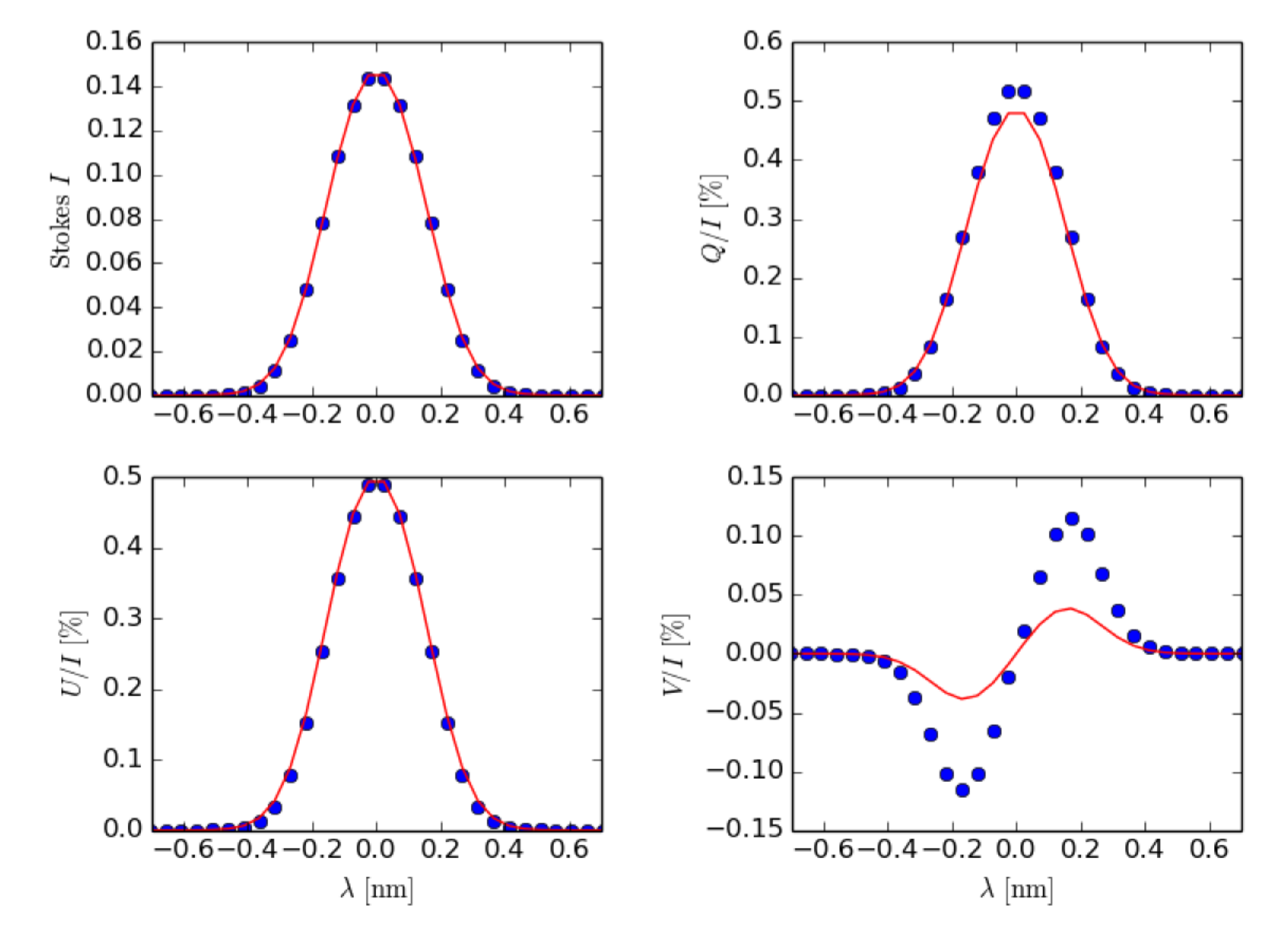}
\includegraphics[width = 0.5\textwidth]{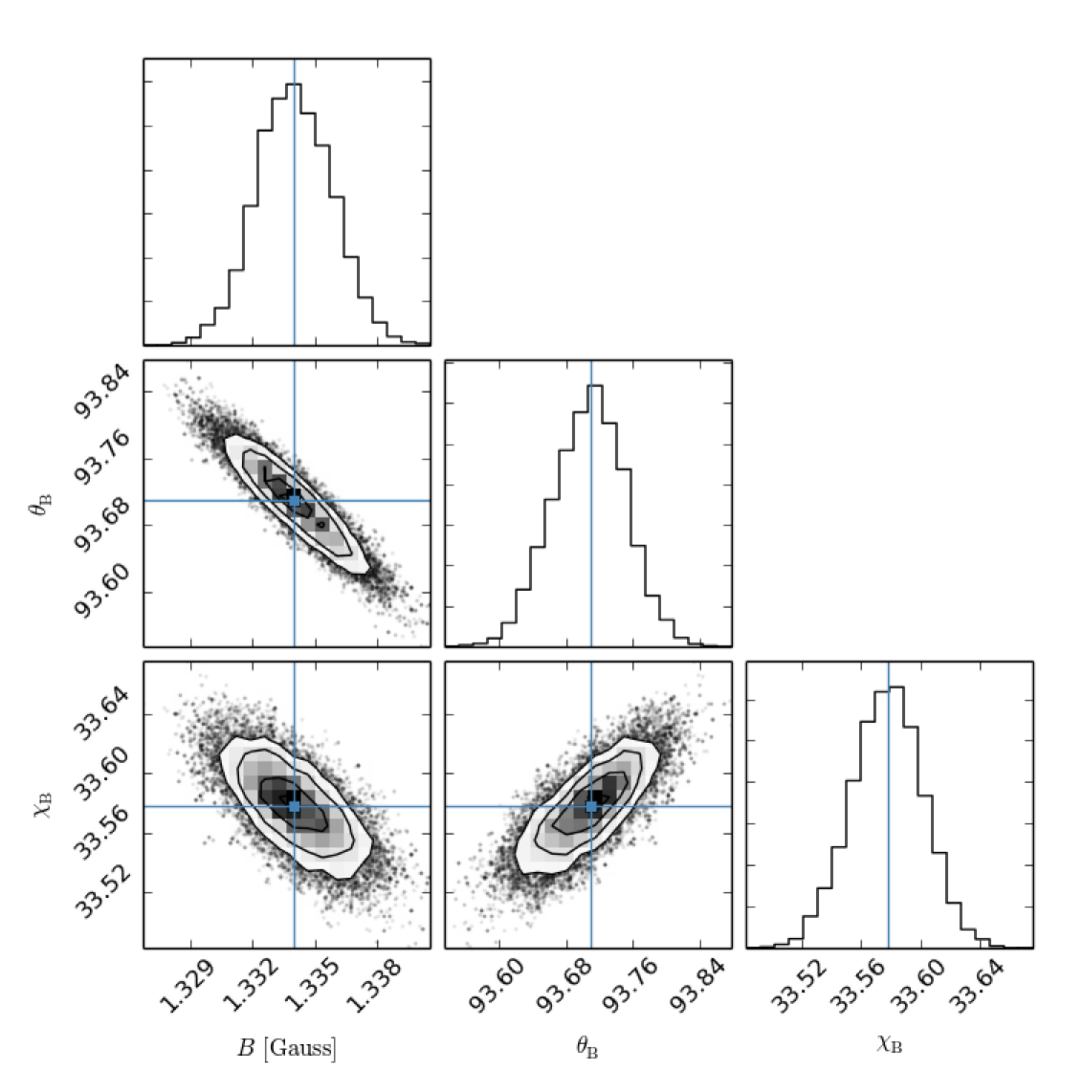}
\caption{Left: Same as in Fig.\,\ref{spectrum_0}, but for pixel at $y=8.5$''. Right: Posterior distributions for the inferred magnetic field vector.}
\label{spectrum_17}
\end{figure}

The above example is not surprising as we expect disagreement on the edges of the slab. In Fig.\,\ref{spectrum_17} we show the best fit for the pixel at $y=8.5\,''$. The fit in the first three Stokes parameters is good, but the inferred parameters (see the corresponding triangle plot) are far from the input values:  we obtain $(B, \theta_B, \chi_B) \approx (1.3\,\rm{G}, 94^o, 33^o)$, i.e.  a significantly weaker and more inclined magnetic field than the input value. We note that the bad fit in Stokes $V$ is not due to the fitting procedure. We  restarted the MCMC sampler several times to find the best fitting profiles. Our interpretation is the following: Because of the multidimensional effects, the linear polarization significantly differs from the value predicted by the 1D model. The fitting procedure then modifies magnetic fields to find the best fit in all  four components of the intensity vector. As the polarization is smallest in Stokes $V$, that is where we find the greatest discrepancy between synthetic observations and the  fit. 

We find it necessary here to discuss briefly the way of visualizing sampled posterior probability. The triangle plot (lower panel of Fig.\,\ref{spectrum_17}) shows the  marginalized posterior probability distributions of parameters (and combinations of parameters) for the three parameters describing the magnetic field vector. For example, the 1D histogram for the values of the magnetic field strength (top panel) is proportional to the following probability density function:
\begin{equation}
 p(B|D) = \int_{\tau, \lambda_0, v_t, \theta_B, \chi_B} p(\hat{M}|D) d\tau\,d\lambda_0\,dv_t\,d\theta_B\,d\chi_B.
\end{equation}
Integration over the remaining parameters is referred to as ``marginalization''. The 2D histograms on the triangle plot are to be understood in a similar manner. These histograms provide us with the shape of posterior probability for the parameters and give us insight into the uncertainties of the inferred parameters. We find it very interesting that the polarized spectra shown in Figs.\,\ref{spectrum_0} and \ref{spectrum_17} can be fit fairly well with a combination of parameters which are significantly different from the input values. Also, the uncertainties for the inferred parameters are very small (these actually depend on the noise values, but that discussion is beyond  the scope of this paper).   

\begin{figure}
\includegraphics[width = 0.5\textwidth]{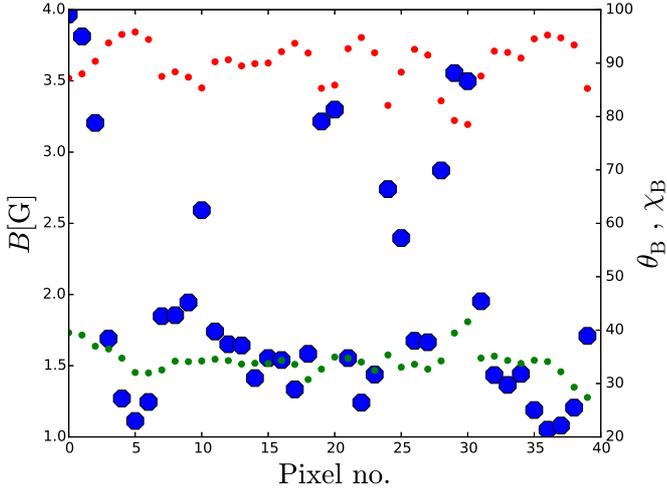}
\caption{Distribution of inferred magnetic field strength (blue), inclination (red), and azimuth (green) over the face of the slab. Polarized spectra used for the inference are given in Fig.\,\ref{example_4}.}
\label{distribution_noiseless}
\end{figure}

These fits are not isolated examples. Actually, for all the inversions, we systematically obtain weaker magnetic fields which are, in general more directed toward the observer than the input values (see Fig.\,\ref{distribution_noiseless}). The interpretation of such a result is not straightforward as it is not easy to intuitively understand the effects of multidimensional radiative transfer. In principle, it is reasonable to expect that inhomogeneity results in increased total polarization. To compensate for this, the 1D inversion procedure invokes a weaker magnetic field (i.e. introduces a smaller amount of depolarization). It also tries to recover the line-of-sight component of the magnetic field as accurately as possible, and thus the inferred magnetic field is directed more toward the observer than the input value. Again, these are the results of the noise-free data. Our next step is to analyze synthetic data that we have computed including velocity fields effects and adding a non-negligible photon noise.    
 
\subsection{Adding velocity fields and photon noise}

\begin{figure*}
\includegraphics[width = \textwidth]{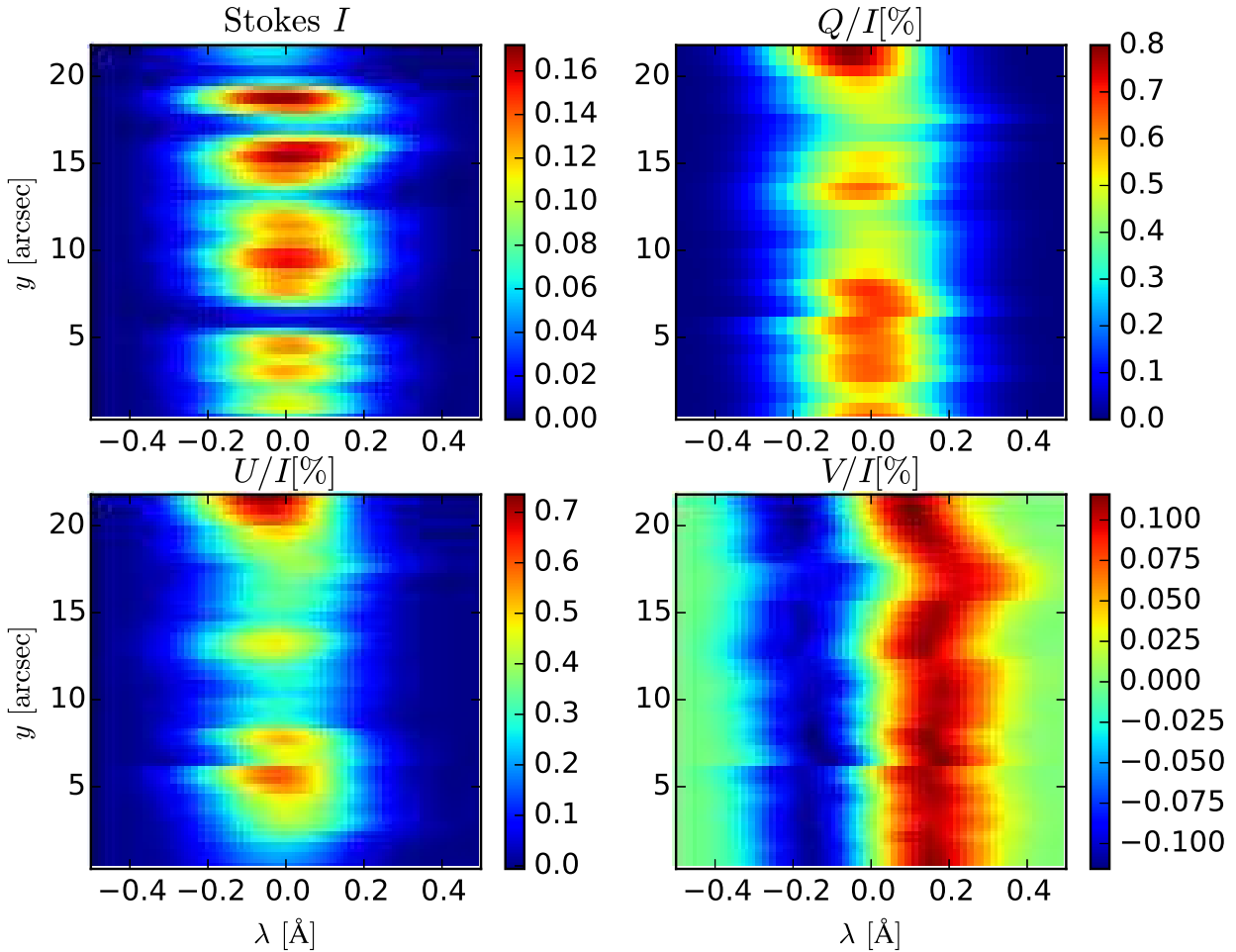}
\caption{Same as Fig.\,\ref{example_4}, except the random realization of threads is different and threads now have systematic velocity (see the text for details).}
\label{example_5}
\end{figure*}

Different kinds of observations and theoretical modeling have shown that prominences (and other atmospheric structures as well) are pervaded by various kinds of velocity fields. These include oscillations \citep{Prom_review_II, HeinzelZapior}, various up- and down-flows, and probably some random but not thermal motions \citep[see the work of]  [where they invoke random thread velocities to explain Ly$\alpha$ intensity distribution]{Gunar12}. In some prominences there are also eruptive plasma velocities directed away from the solar surface, which leave  an imprint on the prominence spectra via a mechanism known as Doppler brightening/darkening \citep{HeinzelRompolt, Labrosse}. Each of these types of velocities can influence the scattering polarization in spectral lines \citep[see the work of][for the influence of dynamics on emergent polarized spectra]{Carlin15}, and probably deserves a paper of its own. Here we restrict ourselves to a very simple case. We assign each thread a velocity equal to the Doppler velocity of the plasma, which is directed randomly. To ensure smooth transition in velocities in the slab, the systematic velocity of each thread falls off following the density of the thread in question. These velocities, in general, shift and broaden the intensity profiles, and can also modify the scattering polarization. An example of  synthetic Stokes spectra  is shown in Fig.\,\ref{example_5}. We note that, in this example, we have a different realization of the thread positions from the one corresponding to Fig.\,\ref{example_4}. 

Before inverting these results, we now also add a random amount of Gaussian noise to them. We assume that the noise at each pixel in Fig.\,\ref{example_5} is proportional to the square root of the appropriate Stokes $I$ and that the noise at the disk center (i.e. for $I=1$) is $3\times10^{-5}$ (which corresponds to approximately $10^9$ counts). While this amount of noise does not influence Stokes $I$ noticeably, the noise is visible in the polarized profiles as they are an order of 2-3 magnitudes smaller. We also take into account the wavelength dependence of the noise and the fact that noise in Stokes $Q$, $U$, and $V$ is larger by a factor $\sqrt{3}$ with respect to noise in Stokes $I$. We comment briefly on the obtained results as the main conclusions do not change with respect to the previous section.

\begin{figure}
\includegraphics[width = 0.5\textwidth]{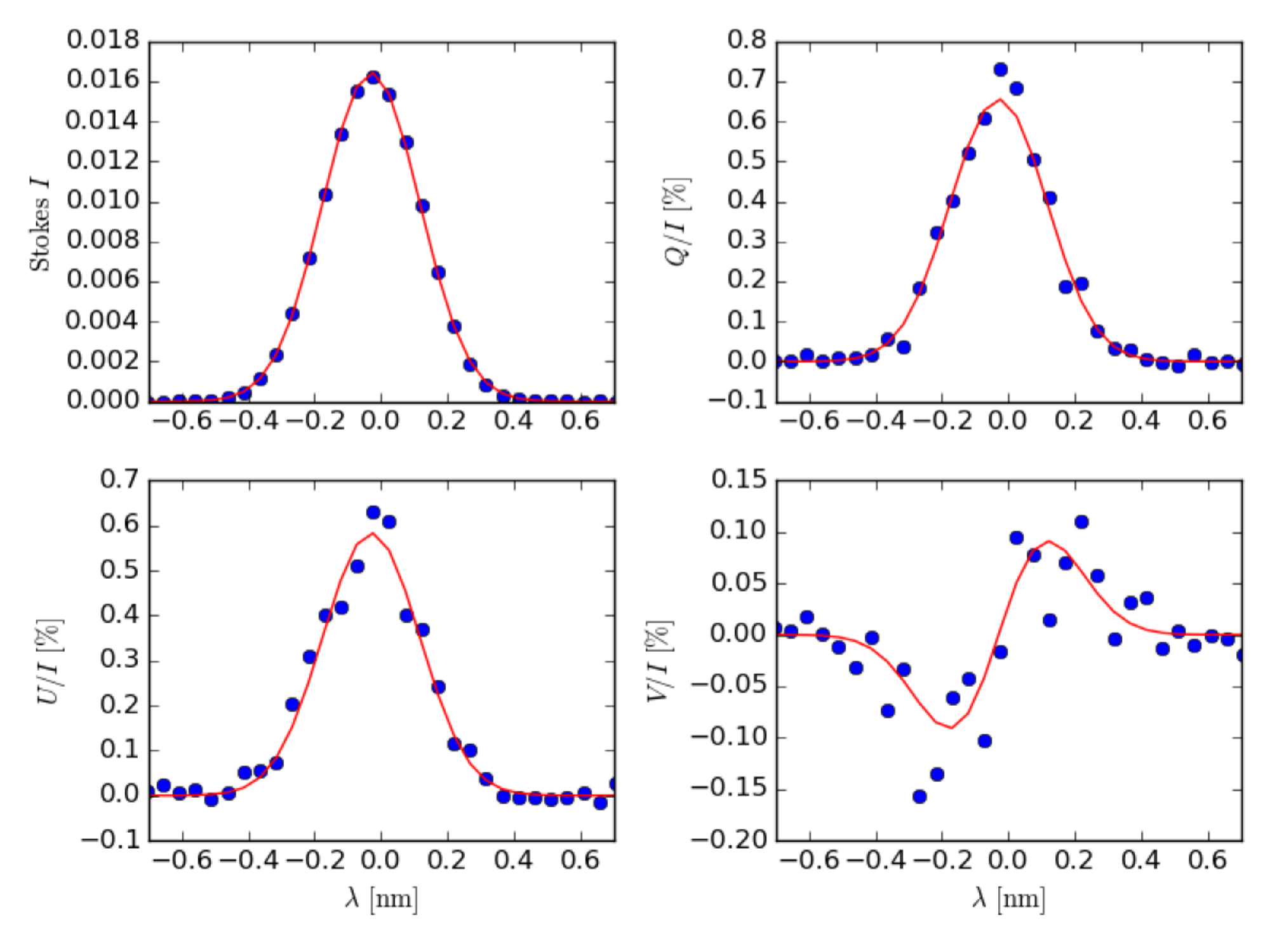}\\
\includegraphics[width = 0.5\textwidth]{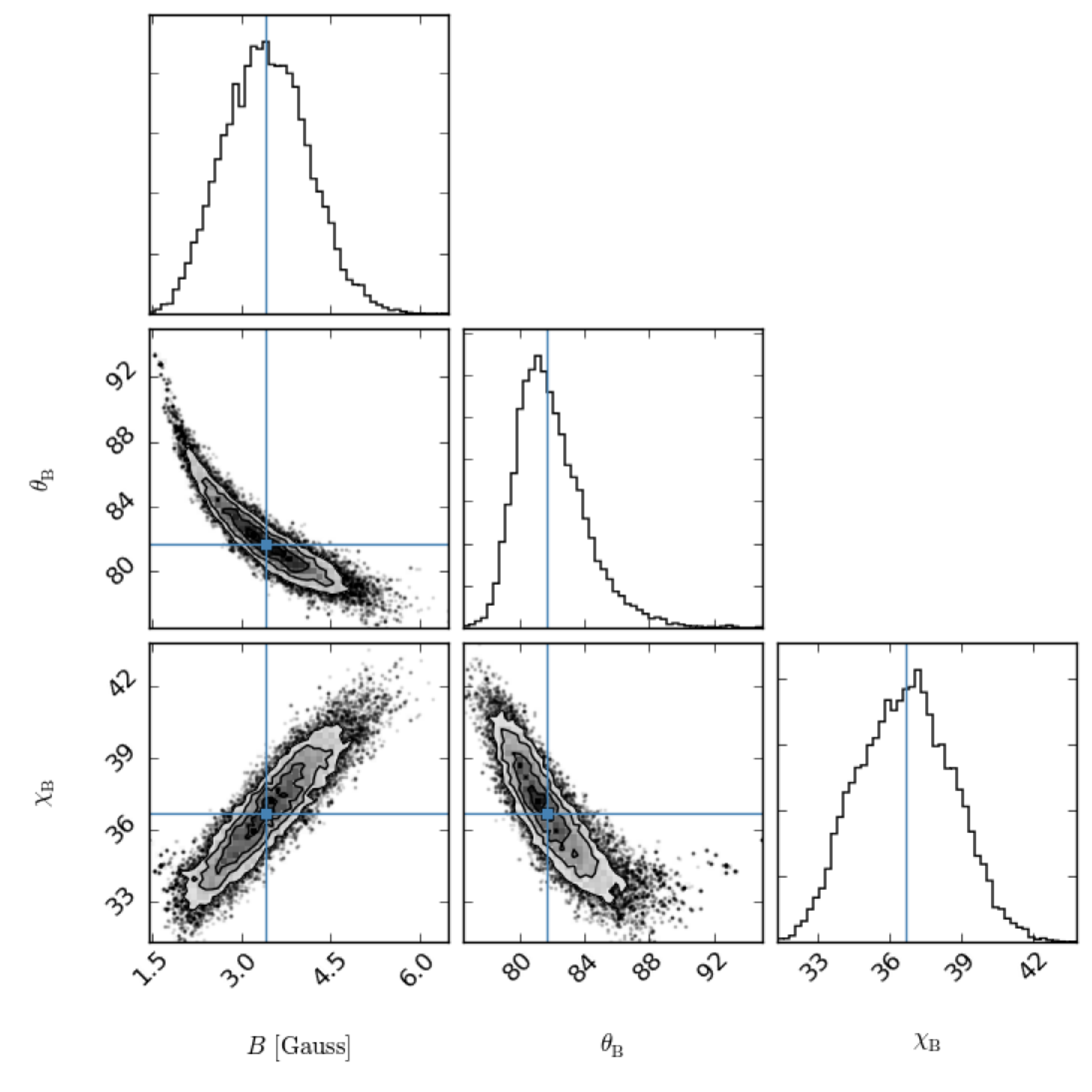}
\caption{Upper panel: Polarized spectrum at location $y=4.5$'' from Fig.\,\ref{example_5}, and the best fit. Lower panel: Corresponding triangle plot.}
\label{spectrum_10}
\end{figure}

After inverting this set of synthetic data we find similar results to those in the previous section. The magnetic field is almost universally weaker and directed more toward the observer. In some of the inverted pixels we have found characteristic, banana-shaped, parameter distributions (Fig.\,\ref{spectrum_10}).  

From the triangle plot given in Fig.\,\ref{spectrum_10}, it is obvious that the posterior distribution of parameters is much broader and asymmetric. The best fit for this particular case is approximately $(B, \theta_B, \chi_B) = (3.5,\rm{G}, 82^o, 37^o)$. The uncertainties are much larger than in the previous section. This is the consequence of the noise. Obviously, a good knowledge of the noise is crucial for the quantitative interpretation of spectropolarimetric observations. This conclusion stands regardless of the main points of this paper and we postpone the discussion of photon noise effects on spectropolarimetric inference to our future investigations.  

\section{Conclusions}

In this paper we have studied the formation of a prototype two-level atom line in a prominence-like structure situated above the solar surface. The structure is finite and inhomogeneous in the $xy$ plane and infinite and homogeneous in the $z$ direction. The line is assumed to have atomic parameters similar to the He\,10830 line. Taking into account the  weak field Zeeman effect and the Hanle effect, we have computed a full Stokes vector emerging from the prominence. We have then applied spatial smearing and binning to obtain synthetic spectra.

We first conclude that the presence of spatial inhomogeneities induces the presence of linear polarization in Stokes $U$ even in the absence of a magnetic field. The magnitude of this polarization is comparable to the magnitude of Stokes $Q$. In the presence of a magnetic field, the interplay of the Hanle effect and of the radiative transfer effects on the resulting Stokes profiles is qualitatively hard to interpret and therefore we have conducted an inversion of the synthetic spectra using an affine-invariant MCMC  scheme. The model used for the inversion is identical to the one used for the forward synthesis except for the fact that it only accounts for 1D radiative transfer. 

The magnetic field vector retrieved by our inversion procedure is more directed toward the observer (i.e. the polar angle is closer to $\pi/2$ and the azimuth is closer to zero) with respect to the
 input magnetic field. Also, the magnitude of the inferred magnetic field is significantly lower ($\approx 2$ versus the original $5$\,gauss). We argue that this is because of the multidimensional and radiative transfer effects which cannot be explained by a simple 1D inversion procedure. We draw a parallel with the real Sun where the 1D inversion schemes are used to interpret line profiles formed in multidimensional environments. We also briefly comment on the influence of systematic velocities and photon noise on the spectra. Moderate, random velocities do not seem to influence the spectrum significantly. The presence of noise, on the other hand, has a strong impact on the inference process. Even the relatively low level of noise we have used here causes broad and asymmetric posterior probabilities of the relevant parameters.
 
It is obvious that prominences are more complex than depicted here. They are, undoubtedly, inhomogeneous along the $z$-axis as well. These are not only inhomogeneities in physical parameters, but also in boundary conditions as lower parts of the prominence ``see'' stronger and less anisotropic radiation. It would be of great importance to study the effects of vertical radiative transfer in this case as the vertical geometrical scales of the problem are similar to the horizontal scales. The most thorough approach would be to synthesize a full Stokes spectrum of realistic spectral lines, coming from a 3D model, and then invert it using state-of-the-art inversion codes (HAZEL, HELIX+). This paper is only the first step in the investigations intended to highlight the problems with using the contemporary inversion techniques in the study of spectral lines formed in non-LTE.

We conclude this paper, which we  hope was instructive, by asking a question: How relevant is this radiative transfer exercise to the interpretation of realistic solar observations? 
It is tempting to ascribe the dominance of the horizontal fields found in the interpretation of prominence observations to these radiative transfer effects. 
In the case where the observed line is optically thin, the line formation is nicely described by a single-scattering approximation, and multidimensional effects are negligible (e.g. the D3 line of helium). 
Inference methods that rely on multi-line observations are also probably immune to these effects. However, for cases where a single, optically thick line is used (e.g. spicules, some prominences), it is possible to encounter line profiles that require detailed radiative transfer modeling. Unfortunately, inversion schemes for the lines formed by scattering which involve multidimensional radiative transfer are (probably) still far away. Detailed, multidimensional NLTE modeling is our best hope to understand the physical conditions in the medium where these lines are formed, both in the chromosphere and in various chromospheric and transition-region structures. 

\begin{acknowledgements}
IM thanks Rafael Manso Sainz for  fruitful discussions on the topics and many useful remarks on the manuscript. IM is thankful to the Observatoire de la C\^{o}te d'Azur for a four-week research grant which made finishing this paper possible. OA and IM both acknowledge the support of the Serbian Ministry of Education and Science through the project 176004 ``Stellar Physics'', as well as financial support from the grant ``Pavle Savi\'{c}.'' 
\end{acknowledgements}

\bibliographystyle{aa}  
\bibliography{ivan_sun} 

\end{document}